\documentclass[aps,pra,showpacs,twocolumn,superscriptaddress]{revtex4}
\usepackage{epsfig}
\begin{document}

\title{
Optimum unambiguous identification of $d$ unknown pure qudit states}
\author{Ulrike Herzog}
\affiliation{Institut f\"ur Physik, Humboldt-Universit\"at Berlin, Newtonstrasse 15, D-12489 Berlin, Germany}
\author{J\'anos A. Bergou}
\affiliation{Department of Physics, Hunter College, City University of New York, 695 Park Avenue,
New York, NY 10021, USA}
\date{\today}

\begin{abstract}
We address the problem of unambiguously identifying the state of a probe qudit with the state of one of $d$
reference qudits. The $d$ reference states are assumed pure and linearly independent but we have no knowledge of
them. The state of the probe qudit is assumed to coincide equally likely with either one of the $d$ unknown
reference states. We derive the optimum measurement strategy that maximizes the success probability of
unambiguous identification and find that the optimum strategy is a generalized measurement. We give both the
measurement operators and the optimum success probability explicitly. Technically, the problem we solve amounts
to the optimum unambiguous discrimination of $d$ known mixed quantum states.
\end{abstract}


\maketitle

\section{Introduction}
It has been shown that $N$ given pure quantum states can be
unambiguously discriminated provided they are linearly independent
\cite{chefles1}, at the expense of a certain fraction of
inconclusive results if the states are nonorthogonal. Unambiguous
state identification is  a variant of the pure-state discrimination
problem. Here we assume that a quantum system of a $d$-dimensional Hilbert space
is prepared in a definite pure state out of a set of
$N$ linearly independent states, but that we do not know the
states in the set. Instead, we are given $N$ reference
systems each being prepared in another one of the $N$ unknown pure
states. Our task is to determine the optimum measurement for
unambiguously identifying the state of the probe system with the
state of one of the reference systems, that is to find the
particular measurement which maximizes the overall probability of a
successful identification. Despite the complete lack of knowledge about
the pure states, their probabilistic identification is possible due
to symmetry properties inherent in the quantum mechanical
description.

Up until now state identification of unknown pure states has only been considered for the case $N=2$. The
problem has been introduced by Bergou and Hillery \cite{BH} for the measurement strategy of optimum unambiguous
identification and was first solved for two unknown qubit states, that is for $d=2$ \cite{BH}. The
identification of two unknown pure states has been shown to be equivalent to the discrimination of two known
mixed quantum states \cite{BBFHH,hayashi1}. The treatment has been extended to the case that the relevant states
are each encoded into a certain number of identical copies of the respective quantum systems
\cite{BBFHH,hayashi1,hayashi2,HeB1,HeB2}, and the measurement strategy of  state identification with minimum
error has been derived, as well \cite{BBFHH,hayashi1}. Moreover, it has been assumed that only one out of two
possible qubit states is unknown and the other is known \cite{BBFHH}. In addition, a number of schemes for
implementing the unambiguous identification of two unknown qubit states have been proposed
\cite{BO,probst,HeBR}. Encoding of the unknown states in quantum systems with Hilbert space dimension $d > 2$
has also been considered for $N=2$, treating both minimum error \cite{hayashi1,IHHH} and optimum unambiguous
identification strategies of two states \cite{hayashi2,HIHH}.

In this paper we focus on the measurement strategy of optimum unambiguous identification. We extend the previous
investigations to allow for an arbitrary number $N$ of linearly independent unknown pure states. Clearly, in
order to encode these states we need quantum systems with dimensionality $d \geq N$. Here we shall consider the
simplest case, assuming that $d=N$, and that the probe qudit is equally likely prepared in any of the $d$
reference states. In Sec. II we establish the connection between the unambiguous identification of $d$ unknown
pure qudit states and the unambiguous discrimination of $d$ known mixed quantum states. Sec. III provides a
general representation  of the detection operators describing the measurement that unambiguously identifies the
unknown pure states. Based on this result, in Sec. IV the optimum structure of the detection operators and thus
the optimum measurement, realizing the maximum success probability, is obtained. Sec. V concludes the paper with
a brief discussion of the results.

\section{Unambiguous identification as mixed-state discrimination}

The total quantum system we are considering consists of one probe qudit, labeled by the index 0, and $d$
reference qudits, labeled by the indices $1, \ldots d$. Let the states $|i\rangle_n$ denote orthonormal
basis vectors for the $n$-th qudit. Then the identity operator in the $d$-dimensional Hilbert space of the
$n$-th qudit is given by
\begin{eqnarray}
I_n= \sum_{i=0}^{d-1} |i\rangle_n  \langle i|_n \qquad(n
=0,1,\ldots,d).
 \label{I-n}
\end{eqnarray}
The projector $P_{0,n}$ onto the two-qudit subspace of dimension
$d^2$ jointly spanned by the eigenstates of the probe qudit and the
$n$-th reference qudit can be written as
\begin{eqnarray}
P_{0,n} = I_0 \otimes  I_n =P_{0,n}^{sym} + P_{0,n}^{as}  \qquad(n
\neq 0),
 \label{P0n}
\end{eqnarray}
where the projectors $P_{0,n}^{as}$ and $P_{0,n}^{sym}$ refer to the
antisymmetric and symmetric part of the joint subspace,
respectively. In the following we shall sometimes omit the symbol
denoting the tensor product. From the explicit expressions for the
antisymmetric and symmetric basis states in the two-qudit subspace
we obtain the representations
\begin{eqnarray}
P_{0,n}^{as} = \sum_{j=1}^{d-1} \sum_{i=0}^{j-1} \frac{|i\rangle_0
|j\rangle_n - |j\rangle_0 |i\rangle_n}{\sqrt{2}} \;\frac{ \langle
i|_0 \langle j|_n -  \langle j|_0 \langle i|_n}{\sqrt{2}}
 \label{P0nasym}
\end{eqnarray}
and
\begin{eqnarray}
\label{P0nsym}
 \lefteqn{P_{0,n}^{sym} =\sum_{i=0}^{d-1} |i\rangle_0 |i\rangle_n \langle i|_0 \langle i|_n}
 \\
&&+\sum_{j=1}^{d-1} \sum_{i=0}^{j-1} \frac{|i\rangle_0 |j\rangle_n + |j\rangle_0 |i\rangle_n}{\sqrt{2}} \;\frac{
\langle i|_0 \langle j|_n +  \langle j|_0 \langle i|_n}{\sqrt{2}}
 \nonumber
\end{eqnarray}
which show that the ranks of $P_{0,n}^{as}$ and $P_{0,n}^{sym}$ are $d(d-1)/2$ and $d(d+1)/2$, respectively.

Let us now assume that the state of the probe qudit coincides with the state of the $n$-th reference qudit, so
that the state of the total $(d+1)$-qudit system can be written as
\begin{equation}
|\Psi_{n}\rangle = |\psi_{n}\rangle |\psi_{1}\rangle \ldots |\psi_{n}\rangle \ldots |\psi_{d}\rangle .
\end{equation}
On the right-hand-side the order of qudits from left to right is probe qudit (zeroth position) followed by the
$d$ reference qudits. If no information is available about the reference states other that they are linearly
independent, the discrimination strategy must be independent of the actual reference states. In order to
emphasize this state independence, we introduce the density operator $|\Psi_{n}\rangle \langle\Psi_{n}|$ and
take its average over the unknown reference states. It will uniformly span the symmetric subspace of the probe
and $n$-th reference qudit and uniformly span the Hilbert spaces of the remaining $d-1$ reference systems. From
symmetry considerations it then follows that, for the case when the probe matches the $n$-th reference state,
the total $(d+1)$-qudit system is  characterized by the average density operator
\begin{eqnarray}
\rho_n = \frac{2}{(d+1)d^d}\;P_{0,n}^{sym} \bigotimes_{i=1\atop
i\neq n}^d I_i \qquad(n = 1,\ldots,d),
 \label{rho-n}
\end{eqnarray}
in analogy to the treatment introduced for the identification of two unknown states \cite{BBFHH,hayashi1}. Here
the pre-factor arises due to normalization, taking into account the ranks of $P_{0,n}^{sym}$ and of the $d-1$
identity operators. Obviously,  since the state of the probe can coincide with any one of the $d$ reference
states, there are $d$ resulting average density operators.

In order to identify the state of the probe qudit, we have to
discriminate among the $d$ different density operators $\rho_n$. A
measurement suitable to accomplish this task can be formally
described with the help of the $(d+1)$ positive detection operators
${\Pi}_1, \ldots {\Pi}_d$ and ${\Pi}_?$ which together span the
identity,
\begin{equation}
\sum_{n=1}^d {\Pi_n}+ \Pi_? = I \equiv I_0 \otimes I_1\otimes \ldots
\otimes I_d. \label{Pi}
\end{equation}
The detection operators have the property that ${\rm
Tr}({\rho_n}{\Pi}_n)$ is the probability of successfully identifying
the density operator as $\rho_n$, while ${\rm Tr}({\rho_n}{\Pi}_m)$
$(m\neq n)$ describes the probability to get an erroneous result and
${\rm Tr}({\rho_{n}}{\Pi}_?)$ is the probability that the
measurement result is inconclusive, i. e. that the attempt at
discrimination fails to give a definite answer
\cite{chefles,springer}. For unambiguous discrimination we have to
require that ${\rm Tr}({\rho_n}{\Pi}_m)=0$ for $m\neq n$, and from
the positivity of the operators $\rho_n$ and $\Pi_m$ it follows that
this requirement can be only met when
\begin{equation}
\label{Pi1} \rho_n \Pi_m = 0 \qquad {\rm if} \quad m \neq n
\end{equation}
\cite{chefles,springer}. When the $d$ mixed states occur with equal
prior probability, given by $1/d$, the overall success probability
of the discrimination measurement takes the form
\begin{eqnarray}
P_{succ}= \frac{1}{d} \sum_{n=1}^d \rm Tr (\rho_n\Pi_n).
\label{P-succ}
\end{eqnarray}
For determining the optimum measurement, we want to find the
detection operators $\Pi_n$ that maximize the success probability
$P_{succ}$ under the constraint that the eigenvalues of the operator
$\Pi_? = I - \sum_{n=1}^d \Pi_n $ have to be non-negative.

\section{General structure of the detection operators}

\subsection{Three unknown qutrit states}

In the following we proceed along the lines of our previous investigations on the unambiguous discrimination of
two mixed quantum states \cite{HB,BFH,herzog} and start by deriving the general structure of the detection
operators $\Pi_n$. It is instructive to consider first only three density operators $\rho_1$, $\rho_2$ and
$\rho_3$ given by Eq. (\ref{rho-n}) with $d=3$ and referring to the unambiguous identification of three unknown
pure qutrit states.

To be specific, let us focus on the detection operator $\Pi_1$ that unambiguously
discriminates the state $\rho_1$ and therefore fulfills the requirement $\Pi_1\rho_2 = \Pi_1\rho_3 =0$.
Clearly, this requirement is only met when the support of $\Pi_1$ is orthogonal to the supports of $\rho_2$
and $\rho_3$, or, in other words, it belongs simultaneously to the kernel of $\rho_2$ and to the kernel of $\rho_3$.
From Eqs. (\ref{P0n}) and (\ref{rho-n}) we can write the projectors onto these two kernels as
\begin{eqnarray}
P_{K_2} = P_{0,2}^{as}\otimes I_1\otimes I_{3},\quad
 \label{SKrho3}
P_{K_3} = P_{0,3}^{as}\otimes I_1\otimes I_{2}.
\end{eqnarray}
Since $I_{1}$ is common to both, the projector $P_{\Pi_1}$ onto the support of $\Pi_1$ can be written as
\begin{eqnarray}
P_{\Pi_1} = I_1 \otimes P^{\prime}_{\Pi_1},
 \label{PPiPrime}
\end{eqnarray}
where $P^{\prime}_{\Pi_1}$ projects onto the subspace spanned by all
states $|\varphi\rangle$ that are linear combinations of the
eigenstates of the operator $P_{0,2}^{as}\otimes I_{3}$, on the one
hand, and also linear combinations of the eigenstates of the
operator $P_{0,3}^{as}\otimes I_{2}$, on the other hand. We denote
the corresponding eigenstates by $|a_l\rangle$ for
 $P_{0,2}^{as}\otimes I_{3}$, and by $|b_l\rangle$ for $P_{0,3}^{as}\otimes
 I_{2}$,
\begin{eqnarray}
2^{-{1}/{2}} (|i\rangle_0|j\rangle_2  - |i\rangle_2|j\rangle_0)\,
|k\rangle_3
 &\rightarrow& |a_l\rangle,
\label{a-l}\\
2^{-1/2} (|i\rangle_0|j\rangle_3 -|i\rangle_3|j\rangle_0
)\,|k\rangle_2 &\rightarrow & |b_l\rangle ,
 \label{b-ll}
\end{eqnarray}
with $i<j$ and $i,j,k=0,1,2$, where $l=1,\ldots,9$ labels the nine
triples $\{i,j,k\}$. For determining $|\varphi\rangle$ we put
\begin{eqnarray}
|\varphi\rangle = \sum_{l=1}^9 a_l |a_l\rangle= \sum_{l=1}^9 b_l
|b_l\rangle,
 \label{s-prime}
\end{eqnarray}
where $a_l$ and $b_l$ are some complex coefficients. Since $a_l=
\sum_{l^\prime} b_{l^\prime}\langle a_l|b_{l^\prime}\rangle$ and
$b_l= \sum_{l^\prime} a_{l^\prime}\langle b_l|a_{l^\prime}\rangle$,
we obtain
\begin{eqnarray}
b_l=\sum_{l^{\prime\prime}=1}^9
b_{l^{\prime\prime}}\left(\sum_{l^{\prime}=1}^9\langle
b_l|a_{l^\prime}\rangle \langle
a_{l^\prime}|b_{l^{\prime\prime}}\rangle\right)=\sum_{l^{\prime\prime}=1}^9
 b_{l^{\prime\prime}}\, B_{ll^{\prime\prime}},
 \label{b-l}
\end{eqnarray}
where  $B_{11}= B_{22}=B_{33}=1/2$,
$B_{12}=B_{21}=B_{23}=B_{32}=-1/4$, $B_{13}=B_{31}=1/4$,
 and $B_{ll^{\prime\prime}}=0$ if
$l,l^{\prime} \geq 4$. Here, Eqs. (\ref{a-l}) and (\ref{b-ll}) have
been used,  with  $l=1$, $l=2$ and $l=3$ standing for $\{i,j,k\}$
equal to $\{1,2,0\}$,$\;\{0,2,1\}$, and $\{0,1,2\}$, respectively.
The only non-trivial solution of the system of equations given by
Eq. (\ref{b-l}) with $l=1,\ldots, 9$ reads $b_1=b_3=-c$, $b_2=c$
with $c$ being an arbitrary constant, while $b_l=0$ if $l\geq 4$.
Upon inserting these values into Eq. (\ref{s-prime}) and applying
Eq. (\ref{b-ll}) we arrive at the normalized state
\begin{eqnarray}
|\varphi_1(3)\rangle =&(-1)^{1} &\frac{1}{\sqrt{6}} \left(
\,|0\rangle_0|1\rangle_2|2\rangle_3
-|0\rangle_0|2\rangle_2|1\rangle_3 \right. \nonumber \\
&+&|2\rangle_0|0\rangle_2|1\rangle_3- |2\rangle_0|1\rangle_2|0\rangle_3  \nonumber \\
&+&|1\rangle_0|2\rangle_2|0\rangle_3-|1\rangle_0|0\rangle_2|2\rangle_3
 \left. \right) ,
 \label{s-prime-1}
\end{eqnarray}
which is the single eigenstate of $P^{\prime}_{\Pi_1}$. In $|\varphi_{n}(d)\rangle$ the
index $n=1$ refers to the fact that it is the eigenstate of this projector
and the argument $d=3$ to the fact that we are dealing with the qutrit case. It should also be noted that $|\varphi_{n}(d)\rangle$ is the completely antisymmetric state of all qudits with the one corresponding to the index omitted.

In a completely analogous way we can represent the projectors on the
other two detection operators as $P_{\Pi_n} = I_n \otimes
|\varphi_n(3) \rangle \langle \varphi_n(3)|$ with $n=2,3$, where
\begin{eqnarray}
|\varphi_2(3)\rangle =&(-1)^{2} &\frac{1}{\sqrt{6}} \left(
\,|0\rangle_0|1\rangle_1|2\rangle_3
-|0\rangle_0|2\rangle_1|1\rangle_3\right.\nonumber\\
&+&|2\rangle_0|0\rangle_1|1\rangle_3- |2\rangle_0|1\rangle_1|0\rangle_3  \nonumber \\
&+&|1\rangle_0|2\rangle_1|0\rangle_3-|1\rangle_0|0\rangle_1|2\rangle_3
 \left. \!\!\right), \qquad
 \label{s-prime-2}
\end{eqnarray}
\begin{eqnarray}
|\varphi_3(3)\rangle =&(-1)^{3}& \frac{1}{\sqrt{6}} \left(
\,|0\rangle_0|1\rangle_1|2\rangle_2-|0\rangle_0|2\rangle_1|1\rangle_2\right.\nonumber\\
&+&|2\rangle_0|0\rangle_1|1\rangle_2- |2\rangle_0|1\rangle_1|0\rangle_2 \nonumber \\
&+&|1\rangle_0|2\rangle_1|0\rangle_2-|1\rangle_0|0\rangle_1|2\rangle_2
 \left.\!\! \right). \qquad
 \label{s-prime-3}
\end{eqnarray}
Thus we obtain for $n=1,2,3$
\begin{equation}
P_{\Pi_n} = I_n\otimes  |\varphi_n(3) \rangle \langle \varphi_n(3)|
=\sum_{k=0}^{2} |\pi^{(k)}_n\rangle \langle \pi^{(k)}_n|,
\label{3-P1}
\end{equation}
where we introduced the normalized states
\begin{eqnarray}
|\pi^{(k)}_n\rangle  =|k\rangle_n \otimes |\varphi_n(3)\rangle
\qquad (k =0,1,2). \label{k-3}
\end{eqnarray}
Since the states $|\pi^{(k)}_n\rangle$  are orthonormal basis states
of the support of $\Pi_n$, it follows that $\Pi_n$ can be
represented as
\begin{eqnarray}
\Pi_n  = \sum_{k,k^{\prime}=0}^{2} \alpha^{(k,k^{\prime})}_{n}
\;|\pi^{(k)}_n\rangle \langle \pi^{(k^{\prime})}_n|\quad(n=1,2,3),
 \label{Pi-1}
\end{eqnarray}
where the coefficients $\alpha^{(k,k^{\prime})}_{n}$ are some
complex constants that have to guarantee the positivity of the
detection operators.   Eq. (\ref{k-3}) yields
$\langle \pi^{(k)}_{m} |\pi^{(k^{\prime})}_n\rangle =
\delta_{k,k^{\prime}}\,\langle \pi^{(k)}_m |\pi^{(k)}_{n}\rangle$
and
\begin{eqnarray}
\langle \pi^{(k)}_{1} |\pi^{(k)}_2\rangle = \langle \pi^{(k)}_{2}
|\pi^{(k)}_3\rangle = \langle \pi^{(k)}_{1} |\pi^{(k)}_3\rangle =
-\frac{1}{3}
 \label{3overlap}
\end{eqnarray}
for $k=0,1,2$, where  Eqs. (\ref{s-prime-1}) - (\ref{s-prime-3})
have been used.

A note is in place here about the choice of the overall signs of the states
$|\varphi_{n}(3)\rangle$. Although they do not affect the projectors
$P_{\Pi_n}$, they lead to sign changes in the overlaps of the $|\pi^{(k^{\prime})}_n\rangle$  states.
For any choice of the overall signs either one or all three of the overlaps are negative. For the purpose of this paper we made the choice given by Eq.
(\ref{k-3}) together with  Eqs. (\ref{s-prime-1}) -
(\ref{s-prime-3}). With this all three
overlaps are negative in Eq. (\ref{3overlap}), reflecting the intrinsic symmetry of the identification problem.

\subsection{d unknown qudit states}

Now we return to the general task of identifying $d$ unknown qudit states or, equivalently, of discriminating
the $d$ density operators given by Eq. (\ref{rho-n}). By the same reasoning that, for $d=3$, led to Eq.
(\ref{PPiPrime}), the projector onto the support of the detection operator $\Pi_n$ is of the form
\begin{eqnarray}
P_{\Pi_n} = I_n \otimes P^{\prime}_{\Pi_n}.
 \label{SPin}
\end{eqnarray}
Here $P^{\prime}_{\Pi_n}$ projects onto the subspace spanned by all states that can be simultaneously written as
linear combinations of the eigenstates of any one of the operators $P_{0,m}^{as} \bigotimes_i I_i$, where $i\neq
m,n$ and $m\neq n$. In the following we shall show that, generalizing Eq. (\ref{3-P1}),
\begin{eqnarray}
P_{\Pi_n} = I_n\otimes  |\varphi_n(d) \rangle \langle \varphi_n(d)|
=\sum_{k=0}^{d-1} |\pi^{(k)}_n\rangle \langle \pi^{(k)}_n|
\label{d-Pn}
\end{eqnarray}
with
\begin{eqnarray}
|\pi^{(k)}_n\rangle  =|k\rangle_n \otimes |\varphi_n(d)\rangle \quad
(k= 0, 1, \ldots d-1),
\label{k}
\end{eqnarray}
where in analogy to Eqs. (\ref{s-prime-1}) - (\ref{s-prime-3})
\begin{eqnarray}
|\varphi_n (d)\rangle  = \frac{(-1)^{n}}{\sqrt{d!}}
\sum_{\sigma}
{\rm sgn} (\sigma)\! \bigotimes_{j=0\atop j\neq
n}^d|\sigma_j\rangle_j.
 \label{pi-n}
\end{eqnarray}
In Eq. (\ref{pi-n}) the sum is taken over all $d!$ permutations
$\sigma$ distributing the excitation numbers $\sigma_{j}=0,1,\ldots, d-1$
over the system of $d$ qudits that we obtain by omitting the $n$-th
reference qudit from the total system of $d+1$ qudits. The
latter are written in fixed order, and ${\rm sgn} (\sigma)$ is the
sign of the permutation. 

Since $\langle \pi^{(k)}_{n} |\pi^{(k^{\prime})}_n\rangle =
\delta_{k,k^{\prime}}$, Eq. (\ref{d-Pn}) is equivalent to the fact
that the detection operators can be written as
\begin{eqnarray}
\Pi_n  = \sum_{k,k^{\prime}=0}^{d-1} \alpha^{(k,k^{\prime})}_{n}
\;|\pi^{(k)}_n\rangle \langle \pi^{(k^{\prime})}_n| \quad
(n=1,\ldots,d),
 \label{Pi-n}
\end{eqnarray}
where the coefficients $\alpha^{(k,k^{\prime})}_{n}$ are complex
constants subject to the constraints $\Pi_n \geq 0$ and $\Pi_? = I -
\sum_{n} \Pi_n \geq 0$. In order to prove Eqs. (\ref{d-Pn}) and
(\ref{Pi-n}), respectively, we first use Eq. (\ref{rho-n}) to
calculate $\rho_m|\pi^{(k)}_n\rangle$, for $m \neq n$, yielding
\begin{eqnarray}
\lefteqn{\rho_m|\pi^{(k)}_n\rangle  \propto }\\
&&  |k\rangle_n \! \sum_{\sigma}{\rm sgn} (\sigma)\!\!\! \bigotimes_{j=1\atop j\neq
m,n}^d\!\!\!|\sigma_j\rangle_j\; P_{0,m}^{sym}|\sigma_0\rangle_0|\sigma_m\rangle_m =0 . \nonumber
 \label{rho1}
\end{eqnarray}
Here the equality sign holds because the contributions of any two terms $P_{0,m}^{sym}|i\rangle_0|j\rangle_m$
and $P_{0,m}^{sym}|j\rangle_0|i\rangle_m$ $(i\neq j)$ cancel due to the opposite sign of the respective
permutations. Hence the requirement for unambiguous discrimination, Eq. (\ref{Pi1}), is met and Eq. (\ref{d-Pn})
is sufficient for the operator $P_{\Pi_n}$ to be the projector onto the support of the detection operator
$\Pi_n$. It remains to be shown that Eq. (\ref{d-Pn}) is also necessary. Because of Eq. (\ref{SPin}) this holds
true if Eq. (\ref{pi-n}) is the only eigenstate of $P^{\prime}_{\Pi_n}$, that is if $P^{\prime}_{\Pi_n}$ is an
operator of rank one. From Eq. (\ref{rho-n}) it follows that for any $m$ with $m\neq n$ we get the
representation $\rho_m = I_n \otimes \rho_m^{\prime}$, where all operators $\rho_m^{\prime}$ together span a
certain Hilbert space ${\cal K}_n$. Because of Eq. (\ref{SPin}) the eigenstates of $P^{\prime}_{\Pi_n}$ also lie
in ${\cal K}_n$, and the requirement $P_{\Pi_n} \rho_m = 0$ is equivalent to $P^{\prime}_{\Pi_n}
\rho_m^{\prime}=0$. This holds for any $m \neq n$. If $P^{\prime}_{\Pi_n}$ is an operator of rank $r$, this
implies that to each of the basis states of the Hilbert space ${\cal K}_n$ jointly spanned by the operators
$\rho_m^{\prime}$  there belong $r$ states that are orthogonal to it and lie also in ${\cal K}_n$. Clearly this
is a contradiction if $r \geq 2$. Thus Eq. (\ref{d-Pn}) with $r=1$ is not only sufficient, but also necessary
for $P_{\Pi_n}$ to be the projector onto the support of $\Pi_n$.

 Before proceeding,  we derive two important
properties of the $(d+1)$-qudit states $|\pi^{(k)}_n\rangle$. First, Eqs. (\ref{k}) and (\ref{pi-n}) show that
for arbitrary $n$ the total number of excitations
 in the $(d+1)$-qudit system described by
$|\pi^{(k)}_n\rangle$  is equal to $k+\sum_{i=0}^{d-1} i$. Hence for any two states with different $k$ the
excitation numbers of at least one of the qudits composing the total system are different and therefore these
states are orthogonal, which leads to
\begin{eqnarray}
\langle \pi^{(k)}_{m} |\pi^{(k^{\prime})}_n\rangle = \delta_{k,k^{\prime}}\,\langle \pi^{(k)}_m
|\pi^{(k)}_{n}\rangle.
 \label{orth1}
\end{eqnarray}
Second, while $\langle \pi^{(k)}_{n} |\pi^{(k)}_n\rangle = 1$, for
$m\neq n$ we find that
\begin{eqnarray}
\label{overlap0}
 \langle \pi_m^{(k)}|\pi^{(k)}_n\rangle = - \frac{1}{d}
 \end{eqnarray}
since the expression
\begin{eqnarray}
\label{overlap0m}
\frac{(-1)^{(n+m)}}{d!}\sum_{\sigma,\sigma^{\prime}} {\rm
sgn}(\sigma){\rm sgn} (\sigma^{\prime})\!\!\!\! \prod_{j=0\atop
j\neq m,n}^d \!\!\! \langle
 k|\sigma_m\rangle_m \langle
\sigma_j^{\prime}|\sigma_j\rangle_j
 \langle \sigma_n^{\prime}|k\rangle_n\! \nonumber
 \end{eqnarray}
is equal to $-1/d$. Here we took into account that the inner
products vanish unless $\sigma_j^{\prime}=\sigma_j$ for $j \neq m,n$
and $\sigma_n^{\prime} =k = \sigma_m$, reducing the double sum to a
single one over $(d-1)!$ permutations which each contribute the
value 1. Eq. (\ref{overlap0}) reflects the intrinsic symmetry of our
identification problem.

\section{The optimum measurement}

Now we apply the general representation of the detection operators
$\Pi_n$, Eq. (\ref{Pi-n}), in order to find the special coefficients
$\alpha^{(k,k^{\prime})}_{n}$ that  determine the optimum operators
$\Pi_n$ which maximize the overall success probability $P_{succ}$,
given by Eq. (\ref{P-succ}). Using
Eqs. (\ref{rho-n}) and (\ref{Pi-n}) a straightforward calculation
shows that
\begin{eqnarray}
{\rm Tr} (\rho_n\Pi_n) &=& \sum_{k,k^{\prime}=0}^{d-1}
\alpha^{(k,k^{\prime})}_{n} \; \langle \pi^{(k)}_n| \rho_n
|\pi^{(k^{\prime})}_n\rangle\\
 & =& \frac{2}{(d+1)d^d}\frac{1}{d!}\sum_{k,k^{\prime}=0}^{d-1}
\alpha^{(k,k^{\prime})}_{n} (d-1)!\,R_{kk^{\prime}}^{(d)},\nonumber
 \label{Tr-n}
\end{eqnarray}
with
\begin{eqnarray}
R_{kk^{\prime}}^{(d)} = \sum_{i=0}^{d-1} \langle i|_0 \langle k|_n
\;P_{0,n}^{sym}|k^{\prime}\rangle_n|i\rangle_0=\frac{d+1}{2}\,\delta_{k,k^{\prime}},
\label{Rkk}
\end{eqnarray}
where   Eq. (\ref{P0nsym}) has been applied in Eq. (\ref{Rkk}).
 Thus, we arrive at
\begin{eqnarray}
P_{succ} = \frac{1}{d} \sum_{n=1}^d {\rm Tr} (\rho_n\Pi_n)=
\frac{1}{d^{d+2}} \sum_{k=0}^{d-1} \left( \sum_{n=1}^d
\alpha_n^{(k)} \right),
 \label{PSucc}
\end{eqnarray}
where $\alpha^{(k)}_n \equiv \alpha_n^{(k,k)}$. In order to maximize
 $P_{succ}$ we need to determine the largest  values
$\alpha_n^{(k)}$ which are still in accordance with the constraint
$\Pi_? = I - \sum_{n} \Pi_n \geq 0$, or $\sum_{n} \Pi_n \leq I$,
respectively, where the operators $\Pi_n$ are given by  Eq.
(\ref{Pi-n}). Since $P_{succ}$ does not depend on the non-diagonal
elements $\alpha_n^{(k,k^{\prime})}$  $(k \neq k^{\prime})$, for
maximizing $P_{succ}$ on the given constraint we have to put
\begin{eqnarray}
\alpha^{(k,k^{\prime})}_n = \alpha^{(k)}_n \delta_{k,k^{\prime}}.
 \label{alpha}
\end{eqnarray}
The constraint then reduces to
\begin{eqnarray}
 \sum_{n=1}^d \Pi_n =   \sum_{k=0}^{d-1} \left(\sum_{n=1}^d \alpha_{n}^{(k)}
|\pi^{(k)}_n\rangle \langle \pi^{(k)}_{n}|\right) \leq I,
 \label{orth}
\end{eqnarray}
where from Eq. (\ref{orth1}) it follows that for different values of
$k$ the operators within the bracket in Eq. (\ref{orth}) act in
orthogonal subspaces. Maximizing the success probability given by
Eq. (\ref{PSucc}) therefore amounts to solving $d$ independent
maximization problems in the $d$ orthogonal subspaces belonging to
different values of $k$. Moreover, since according to Eq.
(\ref{overlap0}) the mutual overlaps $\langle
\pi_m^{(k)}|\pi^{(k)}_n\rangle$ do not depend on $k$, all these
optimization problems are mathematically identical and the index $k$
can be omitted. Therefore our task is to maximize $\sum_{n=1}^d
\alpha_n\equiv S$ with
\begin{eqnarray}
\label{constraint} \sum_{n=1}^d \alpha_{n} |\pi_n\rangle
\langle \pi_{n}| \equiv \Pi_S\leq I,\qquad\\
 \label{overlap1}
 \langle \pi_n |\pi_n\rangle =1,\quad \langle \pi_m |\pi_n\rangle = -\frac{1}{d}
\quad(m\neq n).
 \end{eqnarray}
For this purpose we make use of the method developed by Chefles and
Barnett \cite{chefles2} for the optimum unambiguous discrimination
of linearly independent symmetric pure states.  We first observe
that $d$ state vectors, with mutual overlaps given by
Eq. (\ref{overlap1}), can be represented as
\begin{eqnarray}
|\pi_{n}\rangle = \frac{1}{d}\,|u_0\rangle +
\frac{\sqrt{d+1}}{d}\;\sum_{l=1}^{d-1} {\rm exp}\left({2\pi
i \frac{n}{d}\,l}\right) |u_l\rangle, \nonumber\\
\label{symm}
\end{eqnarray}
where the states $\{|u_l\rangle\}$ with $\langle
u_l|u_{l^\prime}\rangle\ =\delta _{ll^\prime}$ denote some
orthonormal basis in the $d$-dimensional subspace spanned by the set
of states $\{|\pi_n\rangle\}$. Indeed, from Eq. (\ref{symm}) we get
\begin{eqnarray}
\langle \pi_m|\pi_{n}\rangle = -\frac{1}{d} +
\frac{{d+1}}{d^2}\sum_{l=0}^{d-1} \left[ {\rm exp}\left({2\pi i
\frac{n-m}{d}}\right)\right]^l,
 \label{symm1}
\end{eqnarray}
and Eq. (\ref{overlap1}) is immediately recovered using the sum rule for a geometric series. With the help of
Eq. (\ref{symm}) it is easy to check that the states $\{|\pi_n\rangle\}$ belong to the class of symmetric states
that are covariant with respect to the unitary operator $U = \sum_{l=0}^{d-1} {\rm exp}\left({2\pi i
\frac{l}{d}}\right) |u_l\rangle \langle u_l|$ and transform according to $U|\pi_{d}\rangle=|\pi_{1}\rangle$ and
$U|\pi_{n}\rangle=|\pi_{n+1}\rangle$ for $n=1,\dots d-1$. It has been shown \cite{chefles2} that then there
exists an optimum operator $\Pi_S^{opt}$ maximizing $S$ and possessing the same symmetry with respect to $U$,
leading to $\alpha_1=\dots=\alpha_n\equiv \alpha$. The constraint given by Eq. (\ref{constraint}) thus takes the
form
\begin{eqnarray}
 \alpha \sum_{n=1}^d  |\pi_n\rangle \langle \pi_{n}|\leq I =
\sum_{l=0}^{d-1}|u_l\rangle \langle u_l|.
 \label{symm4}
\end{eqnarray}
On the other hand, from Eq. (\ref{symm}) we obtain
\begin{eqnarray}
 \sum_{n=1}^{d}|\pi_n\rangle \langle \pi_{n}|= |u_0\rangle \langle u_0|
 +  \frac{d+1}{d} \sum_{l=1}^{d-1}|u_l\rangle \langle u_l|,
 \label{symm5}
\end{eqnarray}
where the relation
 $\sum_{n=1}^{d}
{\rm exp}[2\pi i  (l-l^{\prime})n/d ]
 =d\;\delta_{ll^{\prime} }$
has been taken into account. Inserting Eq. (\ref{symm5}) into Eq.
(\ref{symm4}) we find immediately that the largest value of
$\alpha$, allowed by the constraint, is given by
$\alpha^{opt}=d/(d+1)$. With $\alpha_n^{(k)} = \alpha^{opt}$, Eqs.
(\ref{PSucc}), (\ref{alpha}) and (\ref{Pi-n}) yield both the maximum
success probability for unambiguously identifying $d$ unknown pure
qudit states,
\begin{eqnarray}
P_{succ}^{opt} =  \frac{1}{(d+1)d^{d-1}},
 \label{Psucc-opt}
\end{eqnarray}
and the optimum detection operators, with $|\pi^{(k)}_n\rangle $ given by Eq. (\ref{k}),
\begin{eqnarray}
\Pi_n^{opt}  = \frac{d}{d+1}\sum_{k=0}^{d-1} \;|\pi^{(k)}_n\rangle
\langle \pi^{(k)}_n| \quad (n=1,\ldots,d).
 \label{Pi-n-opt}
\end{eqnarray}

Equations (\ref{Psucc-opt}) and (\ref{Pi-n-opt}) represent the main results of this paper, generalizing previous
results from two qubits to $d$ qudits. For $d=2$ Eq. (\ref{Psucc-opt}) reproduces the success probability 1/6
obtained previously \cite{BH,BBFHH} for the optimum unambiguous identification of two unknown pure qubit states.
The optimum detection operators $\Pi_n^{opt}$ are not projectors and, therefore, the optimum measurement
strategy is a generalized measurement.

\section{Conclusions}

In this paper we derived the optimum measurement for the unambiguous identification of $d$ unknown pure qudit
states. Our treatment is based on  Eqs. (\ref{orth1}) and (\ref{alpha}), resulting from the special structure of
the density operators to be discriminated, and reduces the optimization problem to $d$ independent
maximization problems in orthogonal subspaces of dimension $d$.  The procedure is analogous to determining the
optimum measurement for the discrimination of two mixed quantum states in cases where the optimization reduces to
maximization problems in orthogonal two-dimensional subspaces \cite{HB,BFH,herzog}.
It should be noted in this context that the correct detection operators have previously been derived by
Zhang {\it el al.} using a slightly different method and without explicitly working out the optimal success
and failure probabilities \cite{zhang}.

Quantum systems with $d > 2$ have been proposed as carriers of quantum information in various
contexts like e. g. quantum cryptography \cite{bruss}, and methods for realizing general linear transformations
on single-photon qudits have been theoretically described \cite{HeBW}. Moreover, the manipulation of biphotonic
qutrits \cite{lanyon} and ququarts \cite{baek} has been experimentally demonstrated. Apart from possible
applications in quantum information, our investigations are also of interest with respect to the theory of
optimum unambiguous discrimination of more than two mixed quantum states \cite{eldar,feng}, where no examples
of explicit solutions have been given before.

\begin{acknowledgments}
We would like to thank Michal Sedl\'ak (Slovak Academy of   Sciences, Bratislava) for generously communicating
related  investigations leading to the hypothesis $ \alpha^{opt}=d/(d+1)$ and, in a second communication,
pointing us to Ref. \cite{zhang}. His hypothesis  prompted us to discover a sign  error in the previous (and
printed) version of our paper, in the  equations corresponding to Eqs.  (\ref{3overlap}) and (\ref{overlap0}) of
the present paper. This  error propagated and led to replacing  $(2d-1)$ with $(d+1)$ in the  final results,
Eqs. (\ref{Psucc-opt}) and  (\ref{Pi-n-opt}).  All  errors are corrected here.
\end{acknowledgments}


\begin{thebibliography}{99}

\bibitem{chefles1} A. Chefles, Phys. Lett. A 239, 339 (1998).

\bibitem{BH}\ J. A. Bergou and M.\ Hillery, \prl {\bf 94}, 160501 (2005).

\bibitem{hayashi1}
A. Hayashi, M. Horibe, and T. Hashimoto,
Phys. Rev. A {\bf 72}, 052306 (2005).

\bibitem{BBFHH} J. A. Bergou, V. Bu\v{z}ek, E. Feldman, U. Herzog, and M. Hillery,
Phys. Rev. A {\bf 73}, 062334 (2006).

\bibitem{hayashi2}
A. Hayashi, M. Horibe, and T. Hashimoto, Phys. Rev. A {\bf 73},
012328 (2006).

\bibitem{HeB1} B. He and J. A. Bergou, Phys. Lett. A {\bf 359}, 103 (2006).

\bibitem{HeB2} B. He and J. A. Bergou,  Phys. Rev. A {\bf 75}, 032316
(2007).

\bibitem{BO} J. A. Bergou and M. Orszag,  J. Opt. Soc.
Am. {\bf B 24}, 384 (2007).

\bibitem{probst} S. T. Probst-Schendzielorz, A. Wolf, M.\ Freyberger,
I.\ Jex, B. He, and J. A. Bergou, Phys. Rev. A {\bf 75}, 052116
(2007).

\bibitem{HeBR}  B. He, J. A. Bergou, and Y. Ren,
 Phys. Rev. A {\bf 76}, 032301 (2007).

\bibitem{IHHH}  Y. Ishida, T. Hashimoto, M. Horibe, and A.\ Hayashi, arXiv:0712.2906 (2007).

\bibitem{HIHH} A. Hayashi, Y. Ishida, T. Hashimoto, and M. Horibe, arXiv:0801.0128 (2008).

\bibitem{chefles} A. Chefles,
Contemp. Phys. {\bf 41}, 401 (2000).

\bibitem{springer}
 J. A Bergou, U. Herzog, and M. Hillery, Lect. Notes Phys. 649, 417-465 (Springer, Berlin, 2004).

\bibitem{HB} U. Herzog and J. A. Bergou, \pra {\bf 71},
050301(R) (2005).

\bibitem{BFH}
J. A Bergou, E. Feldman, and M. Hillery, \pra {\bf 73}, 032107
(2006).

\bibitem{herzog} U. Herzog, \pra {\bf 75}, 052309 (2007).

\bibitem{chefles2} A. Chefles and S. M. Barnett, Phys. Lett. A
250, 223 (1998).

\bibitem{bruss} D. Bru{\ss}  and C. Macchiavello, Phys. Rev. Lett. {\bf
88}, 127901 (2002).

\bibitem{HeBW} B. He, J. A. Bergou, and Z. Wang, Phys. Rev. A {\bf 76},
042326 (2007).

\bibitem{lanyon} B. P. Lanyon
{\it et al.},
Phys. Rev. Lett {\bf 100}, 060504 (2008).

\bibitem{baek} S.-Y. Baek
{\it et al.},
arXiv:0804.3327 (2008).

\bibitem{eldar} Y. C. Eldar, M. Stojnic, and B. Hassibi, \pra {\bf 69},
062318 (2004).

\bibitem{feng} Y. Feng, R. Duan, and M. Ying,
Phys. Rev. A {\bf 70}, 012308 (2004).

\bibitem{zhang} C. Zhang, M. Ying, and B. Qiao, \pra {\bf 74}, 042308 (2006).

\end{thebibliography}
\end{document}